\documentclass[final,authoryear,5p,times,twocolumn]{elsarticle}
\usepackage{graphicx}
\usepackage{amssymb}
\usepackage{natbib}

\journal{New Astronomy}

\newcommand*\diff{
   \mathop{}\nobreak
   \mskip-\thinmuskip\nobreak
   \mathrm{d}}
\providecommand{\e}[1]{\ensuremath{\times 10^{#1}}}
\usepackage{widetext}

\begin{document}

\begin{frontmatter}

\title{How to combine correlated data sets--A Bayesian hyperparameter
matrix method}

\author[har,yar]{Yin-Zhe Ma\corref{cor1}}
\author[har]{Aaron Berndsen\corref{cor2}}

\cortext[cor1]{Email Address: mayinzhe@phas.ubc.ca}
\cortext[cor2]{Email Address: berndsen@phas.ubc.ca}

\address[har]{Department of Physics and Astronomy, University of British Columbia, Vancouver, V6T 1Z1, BC Canada.}
\address[yar]{Canadian Institute for Theoretical Astrophysics, Toronto, Canada.}

\begin{abstract}
We construct a ``hyperparameter matrix'' statistical method for
performing the joint analyses of multiple correlated astronomical
data sets, in which the weights of data sets are determined by
their own statistical properties. This method is a generalization
of the hyperparameter method constructed by \cite{Lahav00} and
\cite{Hobson02} which was designed to combine independent data
sets. The advantage of our method is to treat correlations between
multiple data sets and gives appropriate relevant weights of
multiple data sets with mutual correlations. We define a new
``element-wise'' product, which greatly simplifies the likelihood
function with hyperparameter matrix. We rigorously prove the
simplified formula of the joint likelihood and show that it
recovers the original hyperparameter method in the limit of no
covariance between data sets. We then illustrate the method by
applying it to a demonstrative toy model of fitting a straight
line to two sets of data. We show that the hyperparameter matrix
method can detect unaccounted systematic errors or underestimated
errors in the data sets. Additionally, the ratio of Bayes' factors
provides a distinct indicator of the necessity of including
hyperparameters. Our example shows that the likelihood we
construct for joint analyses of correlated data sets can be widely
applied to many astrophysical systems.
\end{abstract}


\end{frontmatter}

\section{Introduction}
\label{hyper_intro}

Due to the fast development of astronomical observations such as
the measurements of the cosmic microwave background temperature
anisotropy (e.g.~{\it WMAP} \citep{Hinshaw12} and {\it Planck}
\citep{Planck16} satellites) and observations of galaxy clustering
(e.g.~6dF \citep{Magoulas12} and SDSS \citep{Nuza13} galaxy
surveys), more and more large-scale data sets are available for
studying a variety of astrophysical systems. It is, therefore, a
common practice in astronomy to combine different data sets to
obtain the joint likelihood for astrophysical parameters of
interest. The standard approach for this joint analysis assumes
that the data sets are independent, therefore the joint likelihood
is simply the product of the likelihood of each data set. The
joint likelihood function can then be used to determine optimal
parameter values and their associated uncertainties. In the
frequentist approach to parameter estimation, this is equivalent
to the weighted sum of the parameter constraints from the
individual data sets, where the weight of each data set is the
inverse variance. Data sets with small errors provide stronger
constraints on the parameters.

There is a long history discussing the appropriate way to combine
observations from different experiments. In the context of
cosmology, the discussion can be traced back to \cite{Godwin87}
and \cite{Press96}, where weight parameters were assigned to
different data sets to obtain joint constraints on the velocity
field and Hubble parameter $H_{0}$. In these approaches, however,
the assignment of weights to data sets with differing systematic
errors was, in some ways, ad-hoc. For instance, if a data set has
large systematic error and is not reliable, it is always assigned
a weight of zero and is effectively excluded from the joint
analysis. On the other hand, a more trustworthy data set can be
assigned a higher relative weighting.

Due to the subjectivity and limitations of this traditional way of
assigning weights to different data sets, \cite{Lahav00} and
\cite{Hobson02} (hereafter HBL02) developed the original
hyperparameter method. This allows the statistical properties of
the data themselves to determine the relative weights of each data
set. In the framework developed by \cite{Lahav00} and HBL02, a set
of hyperparameters is introduced to weight each independent data
set, and the posterior distribution of the model parameters is
recovered by marginalization over the hyperparameters. The
marginalization can be carried out with a brute-force grid
evaluation of the hyperparameters, or it can be explored by using
Monte Carlo methods which directly sample the posterior
distribution. Such possibilities include Markov chain Monte Carlo
(MCMC) algorithms such as Metropolis-Hastings and Simulated
Annealing, or non-MCMC methods such as Nested
Sampling~\citep{skilling}. The application of hyperparameters was
considered for a variety of cases by~HBL02. For instance, if the
error of a data set is underestimated, the direct combination of
data sets (no hyperparameter) results in an underestimated
error-budget, providing unwarranted confidence in the observation
and producing a fake detection of the signal. The hyperparameter
method, however, was shown to detect such a phenomenon and act to
broaden the error-budget, thus recovering the true variance of the
data sets. By using the hyperparameter method, the results of
joint constraints become more robust and reliable. This approach
has also been applied to the joint analysis of the primordial
tensor mode in the cosmic microwave background radiation (CMB)
\citep{Ma10}, the distance indicator calibration
\citep{Erdogdu03}, the study of mass profile in galaxy clusters
\citep{Host11}, and the cosmic peculiar velocity field study
\citep{Ma12}.

Notably, the hyperparameter method established by \cite{Lahav00}
and HBL02 is limited to independent data sets, where ``no
correlation between data sets'' is assumed in the joint analysis.
In the analysis of cosmology and many other astrophysical systems,
the data sets sometimes are correlated. For instance, in the study
of the angular power spectrum of the CMB temperature fluctuations,
the data from the Atacama Cosmology Telescope (ACT), South Pole
Telescope (SPT) and \textit{Planck} satellite share a large range
of multipole moments $\ell$ (see Fig.~1 of \citealt{Cheng13} and
Fig.~11 of \citealt{Planck15}). When combining these observations,
one needs to consider the correlated cosmic variance term since
these data are drawn from a close region of the sky. In addition,
in the study of the cosmic velocity field \citep{Ma12b}, the bulk
flows from different peculiar velocity surveys are drawn from the
same underlying matter distribution so, in principle, a non-zero
correlation term exists between different peculiar velocity
samples. Therefore, a method both using hyperparameter method and
taking into account the correlation between different data sets is
needed in the study of astrophysics. Providing such a method is
the main aim of this paper.

For a clear presentation, we build up our method step-by-step from
the most basic level, explaining the concepts and derivation
process in a pedagogical way. The structure of the paper is as
follows. In Section~\ref{sec:statistics}, we review Bayes' theorem
(Section~\ref{sec:bayes})
 and the standard multivariate Gaussian distribution
(Section~\ref{sec:multi-Gauss}) in the absence of any
hyperparameters. Section~\ref{sec:hyperparameter} provides a
review of the hyperparameter method as proposed in HBL02. In
Section~\ref{sec:hypermatrix} we present the hyperparameter matrix
method, which is the core of the new method proposed in this
paper. We quote the appropriate likelihood function for the
hyperparameter matrix method for correlated data in
Section~\ref{sec:hypermatrix}, leaving its derivation and proofs
of its salient features in \ref{app:posdef}. The proof of the
functional form for the joint likelihood of correlated data sets
makes use of several recondite matrix operations and lemmas. These
are laid out in~\ref{app:hadamard} and~\ref{block_matrix2}, while
the main text simply quotes their results. In
Section~\ref{sec:test}, we apply our method to a straight-line
model while fitting two independent data sets. We vary the
error-budget and systematic errors in each data set to test the
behaviour of the hyperparameter matrix method. In
Section~\ref{sec:improve}, we also discuss the improvement of our
hyperparameter matrix method over the original method proposed by
HBL02. The conclusion and discussion are presented in the last
section.

\section{Statistical method}
\label{sec:statistics}


\subsection{Bayes theorem}
\label{sec:bayes}

Let us suppose that our data set is represented by $D$ and the
parameters of interest are represented by vector $\vec{\theta}$.
Then by Bayes' theorem, the posterior distribution
Pr($\vec{\theta}|D$) is given by
\begin{equation}
\textrm{Pr}(\vec{\theta}|D)=
\frac{\textrm{Pr}(D|\vec{\theta})\textrm{Pr}(\vec{\theta})}{\textrm{Pr}(D)}\,,
\label{eq:bayes1}
\end{equation}
where $\textrm{Pr}(D|\vec{\theta})$ is called the likelihood
function\footnote{Sometimes it is written as $L(\vec{\theta})$,
but here we stick to the notation Pr$(D|\vec{\theta})$.},
$\textrm{Pr}(\vec{\theta})$ is the prior distribution of
parameters and $\textrm{Pr}(D)$ is the Bayesian evidence, an
important quantity for model selection.

Given a data set $D$, let us suppose we have two alternative
models (or hypotheses) for $D$, namely $H_{0}$ and $H_{1}$. One
can calculate the Bayesian evidence for each hypothesis
$H\in\{H_0,\,H_1\}$ as
\begin{equation}
\textrm{Pr}(D|H)=\int
\textrm{Pr}(D|\vec{\theta})\textrm{Pr}(\vec{\theta}) \diff
\vec{\theta}\,, \label{eq:bayes2}
\end{equation}
where the integral is performed over the entire parameter space
$\vec{\theta}$ of each model $H$. Note that the models may have
different sets of parameters. The evidence is an important
quantity in the Bayesian approach to parameter fitting, and it
plays a central role in model selection~\citep{Jeffreys61,Kass95}.
Specifically, if we have no prior preference between models
$H_{1}$ and $H_{0}$, the ratio between two Bayesian evidences
gives a model selection criterion, or Bayes' factor
\begin{equation}
K=\frac{\textrm{Pr}(H_{1}|D)}{\textrm{Pr}(H_{0}|D)}
=\frac{\textrm{Pr}(D|H_{1})}{\textrm{Pr}(D|H_{0})}\,.
\label{eq:bayes3}
\end{equation}

The value of $K$ indicates whether the model $H_{1}$ is favoured
over model $H_{0}$ by data $D$. \cite{Jeffreys61} gave an
empirical scale for interpreting the value of $K$, as listed in
Table~\ref{tab:evidence}. We will use this table as a criterion to
assess the improvement of statistical significance when using the
hyperparameter matrix method.

\begin{table}
\begin{centering}
\begin{tabular}{@{}ll}\hline
 $K$ value & Strength of evidence \\
 \hline
$<1$ & Negative (supporting $H_{0}$) \\ \hline $1$ to $3$ & Weak \\
\hline
$3$ to $10$ & Substantial \\
\hline
$10$ to $30$ & Strong \\
\hline
$30$ to $100$ & Very Strong \\
\hline
$>100$ & Decisive \\
 \hline
\end{tabular}%
\caption{Jeffreys' empirical criterion for strength of evidence
\citep{Jeffreys61}.} \label{tab:evidence}
\end{centering}
\end{table}

\subsection{Multivariate Gaussian distribution}
\label{sec:multi-Gauss} Let us now consider the combination of
multiple data sets, coming from a collection of different surveys
$S$. Each survey provides $n_i$ number of measurements ($D_i$) of
the quantity we are trying
to fit, whose expectation value by our hypothesis is $\mu_i$. For 
each survey $S_i$ we form the data vector $\vec{x}^{S_i}$ with the
following elements
\begin{equation}
x_{j}^{S_i} \equiv D_j - \mu_j\,, \, j\in\{1,\ldots n_i\}\,.
\end{equation}
The data vector is the difference between the observed value and
the expected value, characterizing the error in the measurement.
As such, it is also referred to as the error vector. We combine
the different data sets by forming a total data vector $\vec{x}$
from the individual survey vectors $\vec{x}^{S_i}$
\begin{equation}
\vec{x}=\left(
\begin{array}{c}
\vec{x}^{S_{1}} \\
\vec{x}^{S_{2}} \\
... \\
\vec{x}^{S_{N}}%
\end{array}%
\right)\,,
\end{equation}%
resulting in a vector with dimension
\begin{equation}
\mathrm{dim}(\vec{x})=\sum_{i=1}^{N}n_{i}=N_{t}\,.
\end{equation}%
In the particular case where all of the data sets have the same
number of samples, the individual data vectors $\vec{x}^{S_i}$
have the same dimension $\mathrm{dim}(\vec{x}^{S_i})=n_{i}\equiv
n$ ($i=1,...N$), and $N_{t}=n\cdot N$.

The covariance matrix\footnote{Note, in
Section~\ref{sec:hypermatrix} of this paper we  use $C$ to
represent the covariance matrix with hyperparameters. $\tilde{C}$
is the special case of $C$ evaluated with all hyperparameters set
to unity.} is, generically,
\begin{eqnarray}
\tilde{C} &=& \left\langle \vec{x}\vec{x}^{T}\right\rangle
\nonumber \\
&=&\left(
\begin{array}{cccc}
\left\langle \vec{x}^{S_{1}}\vec{x}^{S_{1}T}\right\rangle &
\left\langle
\vec{x}^{S_{1}}\vec{x}^{S_{2}T}\right\rangle & ... & \left\langle \vec{x}%
^{S_{1}}\vec{x}^{S_{N}T}\right\rangle \\
\left\langle \vec{x}^{S_{2}}\vec{x}^{S_{1}T}\right\rangle &
\left\langle
\vec{x}^{S_{2}}\vec{x}^{S_{2}T}\right\rangle & ... & \left\langle \vec{x}%
^{S_{2}}\vec{x}^{S_{N}T}\right\rangle \\
... & ... & ... & ... \\
\left\langle \vec{x}^{S_{N}}\vec{x}^{S_{1}T}\right\rangle &
\left\langle
\vec{x}^{S_{N}}\vec{x}^{S_{2}T}\right\rangle & ... & \left\langle \vec{x}%
^{S_{N}}\vec{x}^{S_{N}T}\right\rangle%
\end{array}%
\right)  \nonumber \\
&=&\left(
\begin{array}{cccc}
\left( C^{S_{1}}\right)  & \left( C^{S_{1}S_{2}}\right)
 & ... & \left( C^{S_{1}S_{N}}\right)
 \\
(C^{S_{1}S_{2}}) & \left( C^{S_{2}}\right) & ... & \left(
C^{S_{2}S_{N}}\right)  \\
... & ... & ... & ... \\
(C^{S_{1}S_{N}}) & (C^{S_{2}S_{N}}) & ... & \left( C^{S_{N}}\right) %
\end{array}%
\right)\, ,  \label{new_cov1}
\end{eqnarray}%
where each $C^{S_{i}S_{j}}$ is an $N_{i}\times N_{j}$ matrix,
characterizing the auto- or cross-correlation between the vectors
$\vec{x}^{S_{i}}$ and $\vec{x}^{S_{j}}$.

Finally, the $\chi^{2}$ statistic for the combined data vector
$\vec{x}$ is
\begin{equation}
\chi ^{2}=\vec{x}^{T}\tilde{C}^{-1}\vec{x}\,,
\end{equation}%
and the Gaussian likelihood is
\begin{equation}
\textrm{Pr}(D|\vec{\theta})=\frac{1}{(2\pi)^{\frac{N_t}{2}}\sqrt{\det\tilde{C}}}\exp \left( -\frac{1}{2}\vec{x}^{T}\tilde{C}%
^{-1}\vec{x}\right)\,. \label{eq:like1}
\end{equation}%

Equation~(\ref{eq:like1}) is the Gaussian likelihood function of
$\mu_{i}$ ($i=1,...N$) with respect to the data. However, the
likelihood is a multivariate Gaussian in parameter space only if
the $\mu_{i}$ is a linear function of the parameters of interest.
In a more general case, both $\tilde{C}$ and $\mu_{i}$
($i=1,...N$) in Eq.~(\ref{eq:like1}) may have a dependence on the
model parameters $\vec{\theta}$, so the likelihood function
(\ref{eq:like1}) is not Gaussian in parameter space. But this is
not a problem if we evaluate the likelihood function numerically.


Note that when we combine multiple surveys with correlated data as
in Eq.~(\ref{eq:like1}), we give each data set equal weight, and
combine them all together without distinguishing whether some data
set has poorer estimated error or unaccounted systematic errors.
If a data sets' error and systematic effects are properly
accounted for, this method can give an unbiased estimate of the
parameters of interest. However, if errors or systematic errors
exist, the method can give biased results or exaggerated
significance. We provide several such examples in
Section~\ref{sec:test}, and compare with our hyperparameter matrix
method.

\subsection{Combining independent data sets: Original hyperparameter method}
\label{sec:hyperparameter}


The original hyperparameter method, as proposed by \citet{Lahav00}
and HBL02, assumes that different data sets are independent of one
another. That is $C^{S_{i}S_{j}}=\delta_{ij}C^{S_iS_j}$,
$\delta_{ij}$ being the Kronecker-delta, in which case the
covariance matrix becomes block diagonal. ``Hyperparameters''
$\alpha_i$ are introduced as a rescaling of the error vector
\begin{equation}
\vec{x}_i \rightarrow \vec{x}_i/\sqrt{\alpha_i}\,.
\label{eq:hyper_rescale}
\end{equation}
This is equivalent to rescaling the individual blocks, or data
sets, of the covariance matrix
\begin{equation}
C^{S_{i}}\rightarrow \alpha _{i}^{-1}C^{S_{i}},
\end{equation}%
for the $i$th survey.

With the hyperparameter rescaling of
Equation~(\ref{eq:hyper_rescale}) and the assumption of
independent data sets, the total covariance
matrix becomes%
\begin{equation}
C=\left(
\begin{array}{cccc}
\alpha_{1}^{-1}C^{S_{1}} & 0 & ... & 0 \\
0 & \alpha_{2}^{-1}C^{S_{2}} & ... & 0 \\
... & ... & ... & ... \\
0 & 0 & ... & \alpha_{N}^{-1}C^{S_{N}}%
\end{array}%
\right) .  \label{cov4}
\end{equation}%
Since the autocorrelation $C^{S_i}$ is the covariance of the $i$th
data set, the hyperparameters clearly act to re-weight the
internal errors of the survey $S_i$. Exploring different values of
the hyperparameters is equivalent to exploring potential
systematic errors and error estimates of the data set (see our
examples in Sec.~\ref{sec:test}). In this case the total
$\chi^{2}$ for $N$ combined data sets becomes
\begin{equation}
\chi^{2}=\sum_{i=1}^{N}\alpha _{i}\chi _{i}^{2}\,,
\end{equation}%
and the joint likelihood, including hyperparameter and parameters
of interest, becomes
\begin{equation}
\textrm{Pr}(D|\vec{\theta},\vec{\alpha}) =\prod\limits_{i=1}^{N}\frac{\alpha_{i}^{n_{i}/2}}{\sqrt{(2\pi)^{n_i}\det\left(C^{S_{i}}\right)}}%
\exp \left( -\frac{1}{2}\alpha _{i}\chi _{i}^{2}\right)\,.
\label{eq:like-hyper1}
\end{equation}%
Equation~(\ref{eq:like-hyper1}) is obtained in HBL02 (eq.~(32)) as
the general result of a likelihood function with hyperparameters.
By re-deriving it here, we emphasize the assumption of independent
data sets and show the effect of introducing hyperparameters.
Specifically, a large hyperparameter $\alpha_i$ increases the
error-budget of the $i$th data set and reduces its' constraint in
the likelihood function. Conversely, a small hyperparameter
$\alpha_i$ increases the significance of the $i$th data set.

\subsection{Combining correlated data sets: Hyperparameter matrix method}
\label{sec:hypermatrix}

The original hyperparameter method shown in
Section~\ref{sec:hyperparameter} is only for the case where
different data sets do not have cross-correlation terms, i.e. all
off-diagonal matrix entries $C^{S_{i}S_{j}}=0$ for $i \neq j$. In
this section we generalize the hyperparameter formulism to the
case where correlations between individual data sets is
non-negligible, i.e. generalize to the case when $C$ includes
$C^{S_{i}S_{j}}\neq 0$ if $(i\neq j)$.

As before, for each experiment $i$ we introduce a hyperparameter
$\alpha_i$ as a rescaling of the error vector
\begin{equation}
\vec{x_i} \rightarrow \vec{x_i}/\sqrt{\alpha_{i}}\,,
\end{equation}
but we drop the assumption of independent data sets. The (sub)
covariance matrices become
\begin{equation}
C^{S_iS_j} \rightarrow C^{S_iS_j}/\sqrt{\alpha_{i}\alpha_{j}} \,.
\end{equation}
Therefore, for $N$ correlated data sets, the full covariance
matrix with hyperparameters becomes

\begin{widetext}
\begin{eqnarray}
C=\left(
\begin{array}{cccc}
\alpha _{1}^{-1}C^{S_{1}} & (\alpha _{1}\alpha
_{2})^{-1/2}C^{S_{1}S_{2}} & ...
& (\alpha _{1}\alpha _{N})^{-1/2}C^{S_{1}S_{N}} \\
(\alpha _{1}\alpha _{2})^{-1/2}(C^{S_{1}S_{2}})^{T} & \alpha
_{2}^{-1}C^{S_{2}}
& ... & (\alpha _{2}\alpha _{N})^{-1/2}C^{S_{2}S_{N}} \\
... & ... & ... & ... \\
(\alpha _{1}\alpha _{N})^{-1/2}(C^{S_{1}S_{N}})^{T} & (\alpha
_{2}\alpha
_{N})^{-1/2}(C^{S_{2}S_{N}})^{T} & ... & \alpha _{N}^{-1}C^{S_{N}}%
\end{array}%
\right) ,  \label{cov_generalize2}
\end{eqnarray}%
\end{widetext}

where each $C^{S_{i}}(i=1,...N)$ is an $n_{i}\times n_{i}$
symmetric, positive-definite matrix, while $C^{S_{i}S_{j}}$ is an
$n_{i}\times n_{j}$ asymmetric matrix if $n_{i} \neq n_{j}$.

To simplify the matrix calculations in this case, we define an
$N\times N$
\textit{hyperparameter matrix} $P$ with elements $%
P_{ij}=\left( \alpha _{i}\alpha _{j}\right) ^{-1/2}$
($i,j=1,...N$). Thus
\begin{equation}
P=\left(
\begin{array}{cccc}
\alpha _{1}^{-1} & (\alpha _{1}\alpha _{2})^{-1/2} & ... & (\alpha
_{1}\alpha
_{N})^{-1/2} \\
(\alpha _{1}\alpha _{2})^{-1/2} & \alpha _{2}^{-1} & ... & (\alpha
_{2}\alpha
_{N})^{-1/2} \\
... & ... & ... & ... \\
(\alpha _{1}\alpha _{N})^{-1/2} & (\alpha _{2}\alpha _{N})^{-1/2}
& ... & \alpha
_{N}^{-1}%
\end{array}%
\right) .  \label{hyper_matrix}
\end{equation}%

Note that the covariance matrices $C$
(Eq.~(\ref{cov_generalize2})) and $\tilde{C}$
(Eq.~(\ref{new_cov1})) are $N_{\rm t}\times N_{\rm t}$ matrices,
while $P$ is an $N\times N$ matrix for the $N$ data sets, $N\le
N_t$. The relation between $\tilde{C}$, $P$, $\vec{C}$ cannot be
linked by any ordinary matrix product. Here we define a new
``element-wise'' product ``$\odot$'' which multiplies any $N\times
N$ hyperparameter matrix with any $N_{\rm t}\times N_{\rm t}$
covariance matrix, i.e.
\begin{equation}
C=P \odot \tilde{C}\,. \label{eq:element}
\end{equation}

The $\odot$ operation proceeds as follows. We first expand each
hyperparameter $P_{ij}$ to an $n_{i} \times n_{j}$ matrix by
multiplying the $(\alpha_{i}\alpha_{j})^{-1/2}$ value to an
$n_{i}\times n_{j}$ unit matrix $J_{n_{i}n_{j}}$, where all
elements are equal to one\footnote{In order to distinguish the
unit matrix from the identity matrix, we denote this matrix with
$J$ and identity matrix with $I$; see \ref{app:hadamard} for
illustration.}, while keeping the partition of $P_{ij}$ values the
same as the hyperparameter matrix (\ref{hyper_matrix})---this is
equivalent to the Kronecker product. Then we do a Hadamard product
(see \ref{app:hadamard}) for the extended hyperparameter matrix
with the covariance matrix $\tilde{C}$ (Eq.~(\ref{new_cov1})) to
obtain the total covariance matrix $C$
(Eq.~(\ref{cov_generalize2})).

We can now write the likelihood function which includes both
parameters of interest $\vec{\theta}$ and hyperparameter vector
$\vec{\alpha}$ as
\begin{equation}
\textrm{Pr}(D|\vec{\theta},\vec{\alpha})=\frac{1}{(2\pi)^{\frac{N_t}{2}}\sqrt{\det
\left(C(\vec{\alpha})\right)}}\exp \left(
-\frac{1}{2}\vec{x}^{T}C(\vec{\alpha})^{-1}\vec{x}\right)
,\label{eq:like-hyper2}
\end{equation}%
where we indicate, explicitly, the dependence of $C$ on the
hyperparameter vector $\vec{\alpha}$.


Since the values of hyperparameters $\vec{\alpha}$ can take any
values between $0$ and infinity, we might be worried for the
positive definiteness of the covariance matrix $C(\vec{\alpha})$.
Fortunately there are several important properties of the
hyperparameter covariance matrix $C$ that make it positive
definite, and therefore invertible with a positive determinant.
The rigorous proofs of these properties can be found
in~\ref{app:posdef}. Here, we exploit these useful properties to
greatly simplify the generalized $\chi^2$ calculation, so that
Eq.~(\ref{eq:like-hyper2}) can be re-expressed as
\begin{equation}
\textrm{Pr}(D|\vec{\theta},\vec{\alpha})=\left[
\prod\limits_{i=1}^{N}\left(\frac{\alpha_i}{2\pi}\right)^{n_{i}/2}\right]
 \frac{1}{\sqrt{\det \tilde{C}}}\exp \left( -\frac{1}{2}%
\vec{x}^{T}\left( \hat{P}\odot \tilde{C}^{-1}\right)
\vec{x}\right)\,. \label{like4-copy}
\end{equation}%
In the above expression, $\hat{P}$ is the Hadamard inverse of the
hyperparameter matrix $P$, $\tilde{C}^{-1}$ is the inverse matrix
of the correlation matrix (Eq.~(\ref{new_cov1})) without
hyperparameters, and $\odot$ is the ``element-wise'' product.

This form of the likelihood function is a key result of this work,
which is a generalized expression for the joint distribution of
combined, correlated data sets. As a consistency check, we note
that in the case of independent data sets, $\tilde{C}^{-1}$ is
block diagonal, and Eq.~(\ref{like4-copy}) reduces to the original
hyperparameter likelihood function (Eq.~(\ref{eq:like-hyper1})).
As well, in the case of equal weights to all data sets the
hyperparameter matrix $P$ is the unit matrix, and one recovers the
standard multivariate Gaussian distribution, Eq.~(\ref{eq:like1}).

\section{Example of fitting a straight line}
\label{sec:test} Having derived the joint likelihood function for
correlated data sets together with hyperparameters, we now
investigate a simple demonstrative example of fitting data with a
straight line. The goal is to combine two different data sets for
improved constraints on the posterior distribution
Pr($\vec{\theta}|D$). As was shown in~HBL02, the original
hyperparameter method is particularly useful for overcoming the
common problems of inaccurately quoted error bars and the presence
of systematic errors in the measurements. We reproduce the results
of HBL02 here as a validation of our method, and show that the
hyperparameter matrix extension to correlated data sets also
overcomes these problems, and provides a preferred method for
describing the model parameters.

Starting with the assumption that the underlying model for some
process is a straight line with slope $m=1$ and intercept $c=1$
\begin{equation}
y(x) = m\,x+c\,,
\end{equation}
we generated two independent sets of measurements $D_1$ and $D_2$
for the quantity $y$. For each data set five $x$-values were
randomly drawn from a uniform distribution over $(0,1)$, and the
corresponding $y$-values were drawn from a Gaussian distribution
of known variance $\sigma_k$ and mean $\mu_k = m\,x_k + c$. In
this way we know the ``true'' values of the model parameters $m$
and $c$, which we attempt to recover. Following the notation of
Section~\ref{sec:statistics}, the parameters of interest are
$\vec{\theta}=(m,c)$, and the hyperparameters for the two data
sets are $\vec{\alpha}=(\alpha_1, \alpha_2)$.

In this simple case, where the number of data sets, measurements
and parameters is small, one could determine the posterior
distribution (Eq.~(\ref{eq:bayes1})) by evaluating the likelihood
function and prior distributions on a grid. However, this method
scales geometrically with the number of free parameters and
exponentially with the number of grid points, and can quickly
become impractical to evaluate. Instead, in this work the
posterior distributions of the parameters and hyperparameters (if
present) are obtained using Monte-Carlo Markov-Chains (MCMC).
Specifically, we use the default settings in the PyMC~\citep{PYMC}
framework, which uses a Metropolis-Hastings sampling of the prior
distributions. In this way, the marginalized posterior
distributions for each parameter are recovered from the traces of
the MCMC runs, and the evidence integrals are determined from the
trace of the likelihood function~\citep{Kass95,Raftery07}.
\cite{Weinberg10} points out that using the mean or harmonic mean
of the likelihood function can produce spurious results if there
is a lot of variance in the likelihood, but we have checked the
evidence ratios are consistent with his quadrature formulation.

In the following subsections we investigate the behaviour of the
original hyperparameter and hyperparameter matrix method in
comparison to the standard non-hyperparameter method. We consider
three different cases, as listed in
Section~\ref{sec:accurate_err_no_sys}-\ref{sec:inaccurate_err_no_sys}.
The arrangement of these case studies is similar to that of HBL02,
though we add the case of correlated data sets. This facilitates
readers to directly compare the original hyperparameter method and
our hyperparameter matrix method. For ease of comparison, the
Bayes' factor from each of the different cases is presented in
Table~\ref{tab:Bayesfactor}.

In all cases the prior distributions on the slope $m$ and
intercept $c$ are uniform over the interval $(0,2)$, and the prior
for hyperparameters is $\textrm{Pr}(\alpha)=\exp(-\alpha)$ in the
range $(0,10)$ \footnote{The range of values are chosen in order
to give enough sampling space for hyperparameters.}. Recall that a
hyperparameter of unity is equivalent to no additional weighting,
removing the effects of the weights. As such, it is natural to use
prior distributions for the hyperparameters which give a mean of
one, preferring an analysis with no re-weighting (as did in
HBL02). For ${\rm Pr}(\alpha)= \exp(-\alpha)$, it is a properly
normalized prior function ($\int^{\infty}_{0} {\rm Pr}(\alpha)
{\rm d} \alpha=1$) with mean value equal to unity
($\int^{\infty}_{0} {\rm Pr}(\alpha) \alpha {\rm d} \alpha=1$).
Therefore in the following we will adopt such prior function, and
thus confirm that our results of Bayes' factor are consistent with
the values given by HBL02.


The posterior distributions for the parameters of interest are
obtained by $\mathcal{O}(10^5)$ MCMC steps, with a burn-in of 5000. We
denote the non-hyperparameter method as hypothesis $H_0$, with the likelihood
given by Eq.~(\ref{eq:like1}). $H_1$ is reserved for the original
hyperparameter method, appropriate for data sets with no
correlation, and whose likelihood is given by
Eq.~(\ref{eq:like-hyper1}). Finally, we denote the hyperparameter
matrix method as $H_2$, whose likelihood is given by
Eq.~(\ref{like4-copy}), which allows for correlated data sets.

\subsection{Accurate error-bars and no systematic error}
\label{sec:accurate_err_no_sys}
\subsubsection{Independent data sets}
\label{sec:acc_err_nosys_nocorr} 
\begin{figure*}
\includegraphics[width=1.\textwidth]{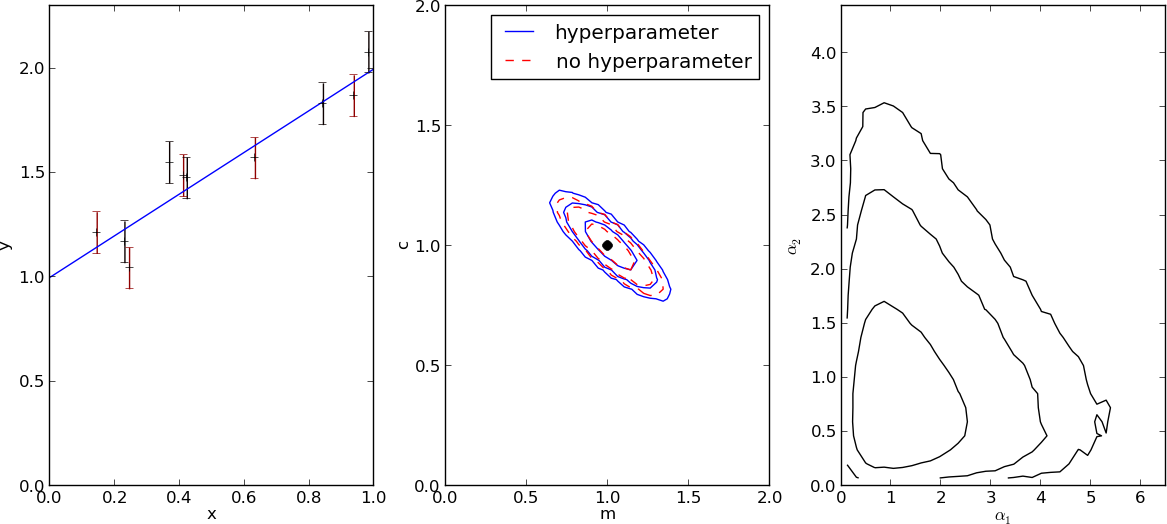}
\caption{\textit{Left}: the two data sets $D_1$ and $D_2$, both
drawn from a Gaussian distribution of mean $\mu = x + 1$ and rms
$\sigma=0.1$. \textit{Middle}: the posterior distributions
Pr$(m,c|D, H_i)$ for the hyperparameter approach of HBL02 ($H_1$,
blue solid lines) and traditional, error-weighted approach ($H_0$,
red dashed lines) approach of parameter estimation. Significance
contours of 68.3\%, 95.4\% and 99.7\% are shown. A black dot
indicates the true values of the model parameters $(m,c)=(1,1)$.
\textit{Right}: the posterior distributions of the hyperparameters
Pr$(\vec{\alpha}|D, H_1)$. Values of unity correspond to no
re-weighting of the data sets, as expected in this case.}
\label{fig:accu_err_nosys_nocorr}
\end{figure*}

In this first case, both data
sets $D_1$ and $D_2$ are drawn from the correct model $m=1$,
$c=1$, with a noise rms of $\sigma_1=\sigma_2=0.1$. From the
experimental side, both data sets are (correctly) assumed to have
an rms of $0.1$ in the likelihood, or $C^{S_1}=C^{S_2}=0.01$.  The
two data sets and the underlying model are shown in the left panel
of Fig.~\ref{fig:accu_err_nosys_nocorr}, with the middle panel
depicting the posterior distributions Pr$(m,c|D,H_i)$ from the
standard non-hyperparameter analysis ($H_0$) and the original
hyperparameter analysis ($H_1$), and the right panel showing the
posterior distributions of the hyperparameters,
Pr($\vec{\alpha}|D, H_1$). In this case both hyperparameters are
consistent with unity, indicating that the hyperparameter method
is not playing an important role in parameter estimation.

Both hypotheses contain the true parameter values $(m,c) = (1,1)$
within the $1\sigma$ confidence level, however the Bayesian
evidence ratio
\begin{equation}
\frac{{\rm Pr}(D|H_1)}{{\rm Pr}(D|H_0)} = 0.61
\end{equation}
indicates that the introduction of hyperparameters is marginally
disfavoured. Here we see one of the powerful results of a Bayesian
approach to combining data sets: the Bayes' factor offers a simple
but distinct method for model selection~\citep{Jeffreys61,
Kass95}. The preference of $H_0$ is not surprising in this case,
since both experiments ``estimated'' the correct variance in the
underlying distributions, $\sigma_1=\sigma_2=0.1$.

\subsubsection{Correlated data sets}
In this section we apply the hyperparameter matrix method, but
where the data sets are (anti) correlated at the 10\% level,
$\rho=C^{S_1S_2}/\sqrt{C^{S_1}C^{S_2}}=-0.1$. As before, both data
sets $D_1$ and $D_2$ are drawn from the correct model $m=1$,
$c=1$, with internal errors of $\sigma_1=\sigma_2=0.1$. In
addition, however, we (correctly) assume a covariance between the
two data sets of $C^{S_1S_2}=-1\e{-3}$, a tenth of the variance of
the two data sets.

\begin{figure*}
\includegraphics[width=1.\textwidth]{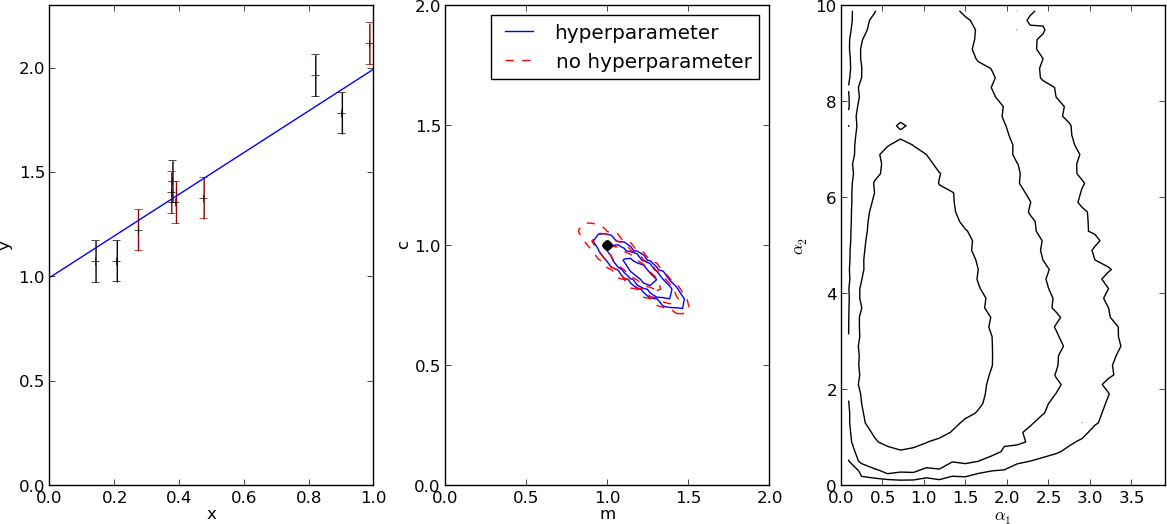}

\caption{{\it Left}: same as Fig.~\ref{fig:accu_err_nosys_nocorr}
but two data sets have a correlation coefficient
$\rho=C^{S_1S_2}/\sqrt{C^{S_{1}}C^{S_{2}}}=-0.1$. {\it Middle}:
the posterior distributions Pr$(m,c|D, H_i)$ for the standard
approach to parameter estimation ($H_0$, red dashed lines) and the
hyperparameter matrix method ($H_2$, blue solid lines). {\it
Right}: the posterior distribution of hyperparameters
Pr$(\vec{\alpha}|D, H_2)$.} \label{fig:accu_err_nosys_corr}
\end{figure*}

With reference to Fig.~\ref{fig:accu_err_nosys_corr} we see,
again, that the correct model parameters are consistent with both
the original and hyperparameter matrix methods, but with an
increased Bayesian evidence factor of 2.56. This is weak support
for the hyperparameter matrix
hypothesis~(Table~\ref{tab:evidence}); however, even in this
simplest of examples, we begin to see that the extended
hyperparameter method provides a better fit to the correlated data
sets than the non-hyperparameter method.

\subsection{Inaccurate error-bars and no systematic error}
\label{sec:inaccurate_err_no_sys}
\subsubsection{Independent data sets}
\label{sec:inaccu_nosys_nocorr} 
\begin{figure*}
\includegraphics[width=1.\textwidth]{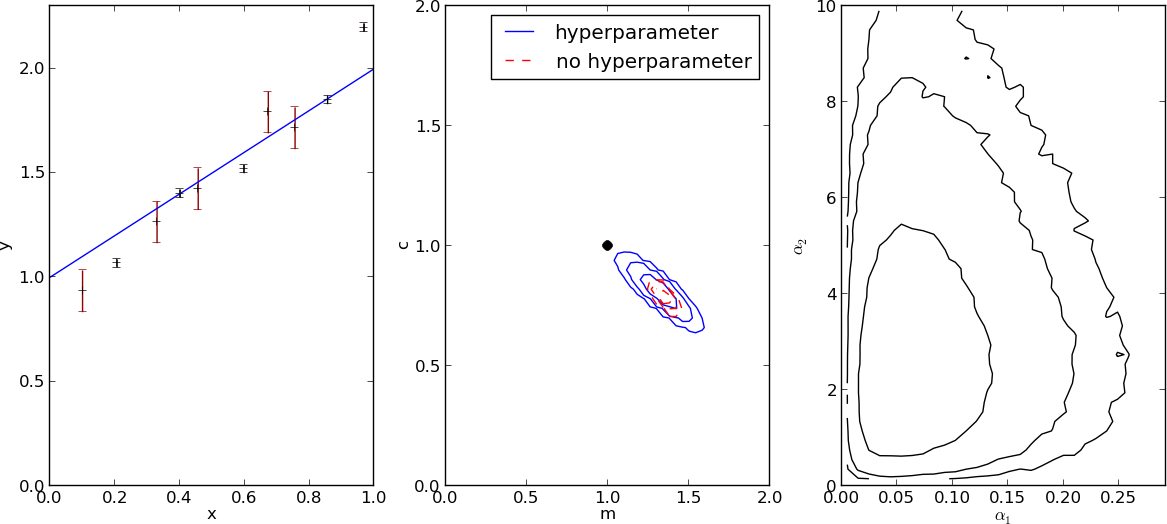}
\caption{{\it Left}: same as Fig.~\ref{fig:accu_err_nosys_nocorr},
but with the reported error bars on data set $D_1$ underestimated
by a factor of 5. {\it Middle}: the posteriors Pr$(m,c|D, H_i)$
corresponding to the standard approach to parameter estimation
($H_0$, red dashed lines) and the hyperparameter method ($H_1$,
blue solid lines). {\it Right}: the posterior distribution of
hyperparameters Pr$(\vec{\alpha}|D, H_1)$.}
\label{fig:inaccu_err_nosys_nocorr}
\end{figure*}

In this case the data sets $D_1$
and $D_2$ are drawn from the same distributions as in
Section~\ref{sec:acc_err_nosys_nocorr}, but in the parameter
estimation procedure, we assume the values of $\sigma_1=0.02$
(underestimated by a factor of $5$) and $\sigma_2=0.1$ in the
likelihood function. In the frequentist approach to parameter
estimation, this underestimation of the noise in $D_1$ would
over-weight its' contribution to the parameter fits. With
reference to Fig.~\ref{fig:inaccu_err_nosys_nocorr}, we see that
the standard non-hyperparameter approach also underestimates the
noise in the parameter fits, so the true value is well outside the
$3\sigma$ confidence level. The original hyperparameter approach,
$H_1$, is consistent with the true parameter values at the
$3.5\sigma$ level and the Bayesian evidence ratio between the two
approaches is $2.5\e{4}$, heavily favouring the hyperparameter
approach. It should be noted that this value is very consistent
with the Bayes' factor obtained by HBL02 in the same case
(sec.~6.2 in HBL02).

\subsubsection{Correlated data}
In this case we draw the two data sets with a correlation
coefficient of $\rho=C^{S_1S_2}/\sqrt{C^{S_1}C^{S_2}}=0.01$, so
the off-diagonal component of the covariance matrix in the joint
likelihood becomes $C^{S_1S_2}=2\e{-5}$. A comparison of the two
posteriors in Fig.~\ref{fig:inaccu_err_nosys_corr} reveals very
different distributions, with the standard non-hyperparameter
method being tightly constrained about the maximum but
inconsistent with the true value. This is a consequence of the
artificially low noise reported for the data set $D_1$. As before,
the Bayesian evidence strongly favours the hyperparameter matrix
approach ($8.3\e{11}$), and a comparison to the evidence ratio for
the case with no correlation between data sets
(Section~\ref{sec:inaccu_nosys_nocorr}) reveals that the
hyperparameter matrix approach deals better with correlated data
sets.

\subsubsection{Interpretation}
\begin{figure*}
\includegraphics[width=1.\textwidth]{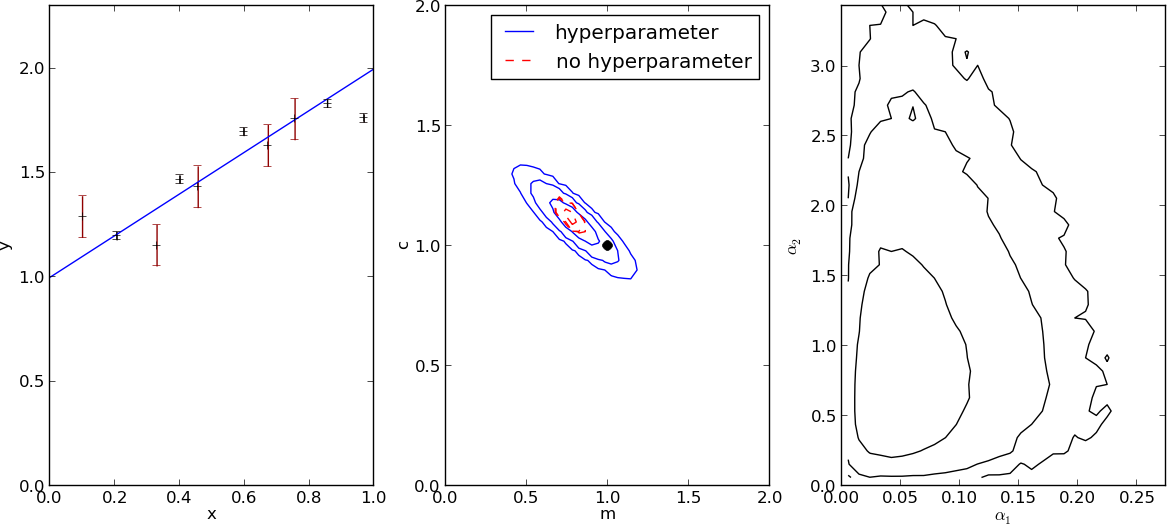}

\caption{{\it Left}: same as
Fig.~\ref{fig:inaccu_err_nosys_nocorr} ($D_{1}$ underestimated
error by a factor of $5$) but with a correlation coefficient
$\rho=C^{S_1S_2}/\sqrt{C^{S_{1}}C^{S_{2}}}=0.01$ between two data
sets. {\it Middle}: the posteriors Pr$(m,c|D, H_i)$ for the
standard ($H_0$, red dashed lines) and hyperparameter matrix
($H_{2}$, blue solid lines) approach of parameter estimation. {\it
Right}: the posterior Pr$(\vec{\alpha}|D, H_2)$ distributions of
the hyperparameters. } \label{fig:inaccu_err_nosys_corr}
\end{figure*}

Let us now understand the values of the hyperparameters. In both
cases of correlated and uncorrelated data sets, the joint
constraint of hyperparameters reveal $\alpha_1\simeq 0.05$, and
$\alpha_2\simeq 1$ (right panels of
Figs.~\ref{fig:inaccu_err_nosys_nocorr} and
\ref{fig:inaccu_err_nosys_corr}). Recalling that the
hyperparameters act to rescale the error vector
$\vec{x}\rightarrow \vec{x}/\sqrt{\alpha}$, and that the error
reported for data set $D_{1}$ was underestimated by a factor of
$5$, we observe that the error recovered by the hyperparameters is
$\sigma_1/\sqrt{\alpha_1}\simeq 0.1$, close to the true value.
Broadly, since $\alpha_1$ is most likely less than $\alpha_2$, the
average effect of the hyperparameters is to reduce the reported
weight of the first data set relative to the second.

To show the importance of the generalized hyperparameter (matrix)
method, we redo the analysis ignoring the data set covariance,
$C^{S_1S_2}=0$, as would be done in the original hyperparameter
method, despite the fact that the data were drawn from a
correlated distribution. A comparison of the evidence for the two
cases gives a Bayes' factor of $1.46$, so recognizing the data
sets having a covariance is a weakly-favoured hypothesis. That is,
the hyperparameter method ($H_1$) is strongly favoured over the
standard joint analysis ($H_0$), and the hyperparameter matrix
method ($H_2$) is weakly favoured over the original hyperparameter
method ($H_{1}$) when errors are mis-reported and correlation
between data set is present. In Sec.~\ref{sec:improve}, we will
sample the correlation strength $\rho$ and show that our
hyperparameter matrix approach provides more reliable fits than
the original method of HBL02.

\subsection{Accurate error-bars with a systematic error}
\label{sec:accurate_err_sys} We have seen that the hyperparameter
matrix approach to combining data sets provides better model
fitting than the standard approach when the reported error bars
differ from the true underlying error. This is true for both the
case of uncorrelated and correlated data sets. In this section we
explore another issue that can corrupt a joint analysis of data
sets: systematic errors. We introduce a systematic error into the
data set $D_1$ by drawing its ``observed'' data from a straight
line with $m=0.5$ and $c=0.5$, while $D_2$ is still drawn from
$m=1$, $c=1$.

\subsubsection{Independent data sets}
\begin{figure*}
\includegraphics[width=1\textwidth]{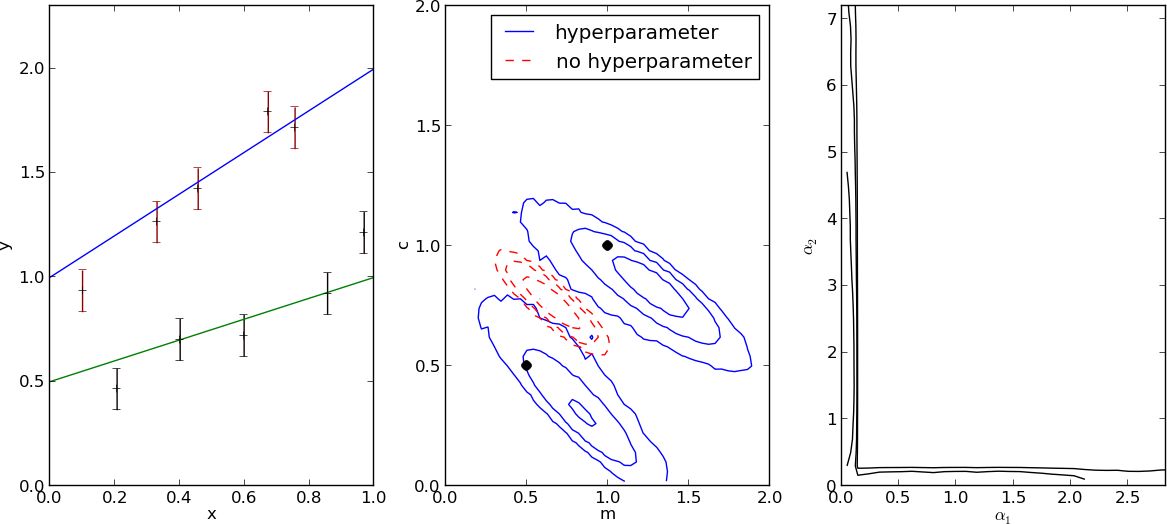}
\caption{{\it Left}: the two data sets $D_1$ and $D_2$ with a
systematic
  difference. One set is drawn from a Gaussian
distribution of mean $\mu = \frac12(x + 1)$ and rms $\sigma=0.1$,
the other from $\mu = x + 1$, $\sigma=0.1$. {\it Middle}: the
posteriors Pr$(m,c|D, H_i)$ corresponding to the standard approach
of parameter estimation ($H_0$, red dashed contours) and the
hyperparameter approach ($H_1$, blue solid contours). {\it Right}:
the posterior distribution of hyperparameters Pr$(\vec{\alpha}|D,
H_1)$.} \label{fig:accu_err_sys_nocorr}
\end{figure*}

Figure~\ref{fig:accu_err_sys_nocorr} shows the two data sets $D_1$
and $D_2$ together with the underlying straight line models from
which they were drawn (left panel). The systematic differences of
the two models are quite apparent, which is reflected in the
posterior distribution of the hyperparameter analysis Pr($m,
c|D,H_1$) (middle panel, blue solid lines).  Pr($m, c|D,H_1$)
clearly indicates a bimodal distribution, recovering the
underlying models $(m,c)=(0.5,0.5)$ and $(m,c)=(1,1)$ at the
$2\sigma$ level. In contrast, the standard non-hyperparameter
approach does not indicate the presence of a systematic difference
between the two data sets, and fails to recover either of the
models with any significance. The result of the joint constraints
clearly reports a wrong parameter space, outside the input values
by more than $3\sigma$ confidence level. The evidence ratio of
$6.1\e{12}$ heavily favours the original hyperparameter approach.

\subsubsection{Correlated data sets}
\begin{figure*}
\includegraphics[width=1.\textwidth]{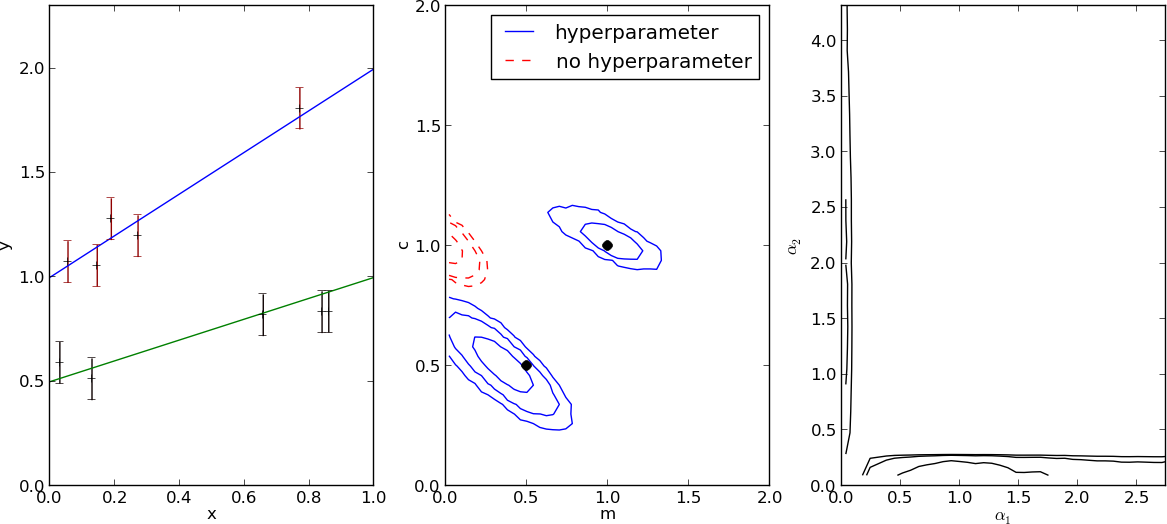}
\caption{{\it Left}: same as Fig.~\ref{fig:accu_err_sys_nocorr}
but two data sets have a correlation coefficient
$\rho=C^{S_1S_2}/\sqrt{C^{S_{1}}C^{S_{2}}}=0.01$. {\it Middle}:
the posteriors Pr$(m,c|D, H_0)$ corresponding to the standard
approach of parameter estimation ($H_0$, red dashed lines) and the
hyperparameter matrix method ($H_2$, blue solid lines). {\it
Right}: the posterior distribution Pr$(\vec{\alpha}|D, H_2)$ of
the hyperparameter matrix method.} \label{fig:accu_err_sys_corr2}
\end{figure*}

In this case, we compare the non-hyperparameter likelihood
analysis ($H_{0}$) with the hyperparameter matrix approach when a
systematic is present in one of the data sets, and there is a
correlation between the two with coefficient $\rho=0.01$ (i.e.
$C^{S_1S_2}=1\e{-4}$). The posterior distributions recovered in
this situation are shown in Fig.~\ref{fig:accu_err_sys_corr2},
with the similar result that the hyperparameter approach reveals a
bimodal distribution, indicating the presence of a systematic
difference in the data sets. The Bayes' factor comparing $H_2$ to
$H_0$ is $1.5\e{15}$.


Right panels in Figures~\ref{fig:accu_err_sys_nocorr} and
\ref{fig:accu_err_sys_corr2} show the marginalized distribution of
hyperparameters $\alpha_{1}$ and $\alpha_{2}$. One can see that
since the two data sets have systematic errors, the constraints on
hyperparameters have two branches. In each branch, one parameter
takes an ordinary value while the other is close to zero. This is
a consequence of the presence of a systematic, since the error is
reduced by ignoring one of the data sets entirely instead of
combining them jointly.

\begin{table}
\begin{centering}
\begin{tabular}{c|c|c|c}
 accurate & systematic & correlated & Bayes'  \\
error bars & error & data sets & Factor \\
 \hline
Y & N & N & 0.6\\
Y & N & Y & 2.6 \\
\hline
N & N & N & 2.5\e{4}\\
N & N & Y & 8.3\e{11}\\
\hline
Y & Y & N & 6.1\e{12}\\
Y & Y & Y & 1.5\e{15} \\
 \hline
\end{tabular}%
\caption{Ratio of Bayes' evidence factors of the hyperparameter
analysis to the standard non-hyperparameter analysis under varying
cases of systematic errors, inaccurate error bars, and correlated
data sets. The last column is Bayes' factor $K$
(Eq.~(\ref{eq:bayes3})). We calculate this $K$ factor with the
original hyperparameter method ($H_{1}$) over standard
non-hyperparameter analysis ($H_{0}$) for uncorrelated data sets,
and the hyperparameter matrix method ($H_{2}$) over standard
Gaussian likelihood analysis ($H_{0}$) for correlated data sets.
Note that throughout the calculation we adopt the exponential
prior on hyperparameters (${\rm Pr}(\alpha)=\exp(-\alpha)$).}
\label{tab:Bayesfactor}
\end{centering}
\end{table}

\subsection{The improvement on the original hyper-parameter method}
\label{sec:improve}

\begin{figure}
\centerline{\includegraphics[width=3.2in]{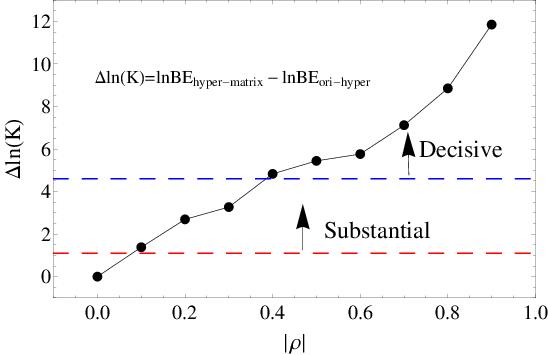}} \caption{The
difference between logarithmic Bayesian evidence (the factor BE is
defined as Eq.~(\ref{eq:bayes2})) as a function of the correlation
strength $\rho$. $\rho$ is sampled from $0$ to $-0.9$ with each
step $-0.1$. $\Delta \ln BE$ is equal to the value of the Bayesian
evidence with our hyperparameter matrix method to consider full
covariance between data sets, minus the value of Bayesian evidence
from the original hyperparameter method (ignore the correlation
between data sets). For the specific experiment, please refer to
Sec.~\ref{sec:improve}.} \label{fig:correlate}
\end{figure}

The hyperparameter matrix method we propose here is the most
general method which can be used to combine arbitrary number of
multi-correlated experimental data. This greatly breaks up the
limitation of the original hyperparameter method (\cite{Lahav00}
and HBL02) which can only deal with multiple independent data
sets. It is always important, to include all of the correlation
information between data sets to obtain correct parameter values
and justify the goodness of fit.

To see the importance of our method, we design an illustrative
experiment to demonstrate this. We generate two data sets with
$N=5$. For each data set, we generate the samples with mean $1.0$
and $0.0$ with Gaussian error $\sigma=0.1$ but correlated between
the two data sets. We take the correlation strength $\rho$ as
$0.0$, $-0.1$, $-0.2$,..., $-0.9$.

Then we use these correlated data sets to do a parameter
estimation. We first use our hyperparameter matrix method, which
considers the full covariance matrix between two data sets. Then
in order to check the behaviour of the original hyperparameter
method, we {\it ignore} the correlation part of the two
experiments and treat them as individual data sets. We calculate
the Bayesian evidence value (Eq.~(\ref{eq:bayes2})) for both
cases, and obtain the difference between the two Bayesian evidence
(BE) values.

In Fig.~\ref{fig:correlate}, we plot the difference between BE
value for our hyperparameter matrix method and for the original
hyperparameter method. First, one can see that when $\rho=0.0$,
the two methods are the same one so $\Delta \ln(K)=0$. But as the
correlation strength increases, the $\Delta \ln(K)$ increases as
well, indicating that the hyperparameter matrix method provides
better and better fits than the original hyperparameter method.
This can be understood as the danger of ignoring correlation
between data sets, since the model becomes inadequate to fit the
data if the correlation is not included. In
Fig.~\ref{fig:correlate}, one can see that if $|\rho|>0.1$, the
Bayes' factor becomes ``Substantial'', and if $|\rho|>0.4$, the
Bayes' factor becomes ``Decisive''. This strongly indicates that
when combining multiple correlated data sets, it is very necessary
to use our hyperparameter matrix method rather than the original
hyperparameter method.

\section{Conclusion}
\label{sec:conclude} In this paper we have reviewed the standard
approach to parameter estimation when there are multiple data
sets. This is an important aspect to most scientific enquiries,
where multiple experiments are attempting to observe the same
quantity. In the context of a Bayesian analysis, the data can also
be used for model selection and tests of the null hypothesis. We
reviewed the original hyperparameter method of~HBL02 for combining
independent data sets, showing how it can overcome inaccurate
error bars and systematic differences between multiple data sets.

Here we developed the hyperparameter matrix method for the case of
correlated data sets, and we have shown that it is a preferred
model to the standard non-hyperparameter approach of parameter
estimation. We rigorously prove that the hyperparameter matrix
likelihood can be greatly simplified and be easily implemented.
From this form of the likelihood, we can recover the simple case
of no hyperparameters where all of the data sets have equal
weights. As well, the original hyperparameter approach is
recovered in the limit of no inter-data set covariance
($C^{S_iS_j}=0$ if $i\neq j$), so our likelihood function provides
a generalized form which covers hyperparameter and
non-hyperparameter analysis, as well as correlated and
uncorrelated data sets.

We test this statistical model by fitting two data sets to a
straight line, and looked at the consequences of mis-reported
error bars, as well as systematic differences between correlated
data sets. In all cases, with the assistance of Bayesian evidence,
we find that the hyperparameter matrix method is heavily favoured
over the traditional joint analysis. By using an illustrative
example to calculate the difference of Bayesian evidence value
between the hyperparameter matrix method, and the original
hyperparameter method, we demonstrate that the Bayes' factor
becomes very substantial (decisive) if $|\rho|$ is greater than
$0.1$ ($0.4$). This suggests that for the case where two
experiments are strongly correlated, our hyperparameter matrix
method is heavily favoured over the original hyperparameter
method.

The method proposed here can be used in a variety of astrophysical
systems. In the context of cosmology, when cosmic variance is a
common component to all large-scale observations, the data sets
drawn from the same underlying density or temperature field will
be correlated to some degree. For instance, in the study of CMB
where multiple data sets drawn from the same region of the sky are
combined (such as \textit{Planck} \citep{Planck16}, \textit{WMAP}
\citep{Hinshaw12}, SPT \citep{Hou12} and ACT \citep{Sievers13}),
it is necessary to consider the correlation between data sets
since they follow the same underlying temperature distribution.
Therefore our method can be an objective metric to quantify the
posterior distribution of cosmological parameters estimated from
the CMB. In addition, in the analysis of the galaxy redshift
surveys for cosmic density and velocity fields, when combining two
surveys data drawn from the similar cosmic volume, the cosmic
variance between different data sets should also be considered as
a part of the total covariance matrix since they all follow the
same underlying matter distribution. In the future survey of
21~cm, if two or more surveys sample the neutral hydrogen in the
same (or close) cosmic volume, the correlation between surveys
should also be considered when combining data sets. In this sense,
our hyperparameter matrix method provides an objective metric to
quantify the probability distribution of the parameters of
interest when multiple data sets are combined.

In summary, when combining correlated data sets, the
hyperparameter matrix method can provide an unbiased and objective
approach that can wisely detect and down-weight any unaccounted
experimental errors or systematic errors, in this way it provides
the most robust and reliable constraints on astrophysical
parameters.

\section{Acknowledgements}

We would like to thank Chris Blake, Andrew Johnson, Douglas Scott
and Jasper Wall for helpful discussions. Y.Z.M. is supported by a
CITA National Fellowship. This research is supported by the
Natural Science and Engineering Research Council of Canada.

\appendix

\section{Theorem: positive-definiteness of the hyperparameter
covariance matrix} \label{app:posdef}

The generalized form of the likelihood function for the
hyperparameter analysis in the presence of correlated data sets
(Eq.~(\ref{eq:like-hyper2})) must satisfy several properties in
order to serve as a probability density function. In particular,
the generalized hyperparameter covariance matrix
$C=P\odot\tilde{C}$ (Eq.~(\ref{cov_generalize2})) must have
positive determinant, and must be invertible.  However, since the
matrix $P$ is a function of the hyperparameters $\vec{\alpha}$
which, in principle, vary from zero to infinity, the positive
definiteness and invertibility of $C$ are not immediately clear.

The following theorem guarantees the feasibility of inverting the
total covariance matrix $C$, and the positive definiteness of the
determinant.

\textbf{Theorem:} The likelihood function of combining $N$
correlated data sets with hyperparameter matrix, i.e.
Eq.~(\ref{eq:like-hyper2}) is equivalent to
\begin{equation}
\textrm{Pr}(D|\vec{\theta},\vec{\alpha})=\left[
\prod\limits_{i=1}^{N}\left( \frac{\alpha
_{i}}{2\pi}\right)^{n_{i}/2}\right] \frac{1}{\sqrt{\det \tilde{C}}}\exp \left( -\frac{1}{2}%
\vec{x}^{T}\left( \hat{P}\odot \tilde{C}^{-1}\right)
\vec{x}\right) , \label{like4}
\end{equation}%
where $n_{i}$ is the dimension of the $i$th data set, $\tilde{C}$
is the covariance matrix between $N$ data sets without the
inclusion of hyperparameter (Eq.~(\ref{new_cov1})), $\odot$ is the
element-wise product (same as Eq.~(\ref{eq:element})), and
$\hat{P}$ is the ``Hadamard inverse'' of the $P$ matrix
(see~\ref{app:hadamard}).

We first prove the inverse relation,
\begin{eqnarray}
C^{-1} &\equiv &\left(
P\odot \tilde{C}\right) ^{-1}  \nonumber \\
&=&\hat{P}\odot \tilde{C}^{-1}.  \label{inverse_new1}
\end{eqnarray}%

\textbf{Proof.}~(1) Let us multiply matrices $\left(P \odot
\tilde{C}\right)$ and $\left(\hat{P}\odot \tilde{C}^{-1} \right)$,
then take the block element $(i,j)$ of the matrix, i.e. ``$i$,
$j$, $k$'' are the block element which can take any value between
($1,...,N$)
\begin{eqnarray}
&&\left[ \left( P\odot \tilde{C}\right) \left( \hat{P}\odot \tilde{C}%
^{-1}\right) \right] _{ij}  \nonumber \\
&=&\sum_{k}\left( P\odot \tilde{C}\right) _{ik}\left( \hat{P}\odot \tilde{C}%
^{-1}\right) _{kj}  \nonumber \\
&=&\sum_{k}\left( \tilde{C}_{ik}\ast \left( \alpha _{i}\alpha
_{k}\right) ^{-1/2}\right) \left( \tilde{C}_{kj}^{-1}\ast \left(
\alpha _{k}\alpha
_{j}\right) ^{1/2}\right)  \nonumber \\
&=&\sum_{k}\left( \tilde{C}_{ik}\tilde{C}_{kj}^{-1}\right) \left(
\alpha
_{j}/\alpha _{i}\right) ^{1/2}  \nonumber \\
&=&(\delta _{ij})I_{n_{i}\times n_{j}}\left( \alpha _{j}/\alpha
_{i}\right)
^{1/2}  \nonumber \\
&=&(\delta _{ij})I_{n_{i}\times n_{i}}, \label{eq:proof1}
\end{eqnarray}%
where in the second step, we use the property of block matrix
product. The final line of Eq.~(\ref{eq:proof1}) indicates that,
only if $i=j$, the product is an $n_{i} \times n_{i}$ identity
matrix, otherwise it is all zeros. Thus we prove the inverse
relation (Eq.~(\ref{inverse_new1})). $\Box$

Next, let us prove the determinant relation%
\begin{equation}
\det(C)=\det \left( P\odot \tilde{C}\right) =\det \tilde{C}\ast
\left( \prod\limits_{i=1}^{N}\alpha _{i}^{-n_{i}}\right) ,
\label{detC_hyper1}
\end{equation}%
where $C$ is given by Eq.~(\ref{cov_generalize2}), $\tilde{C}$ is
given by (Eq.~(\ref{new_cov1})) and $n_{i} $ is the dimension of
the $i$th block matrix.

\textbf{Proof.}~(2) In \ref{block_matrix2}, we have proved that a
matrix of type $\tilde{C}$ (\ref{new_cov1}) follows the
determinant Eqs.~(\ref{det1})-(\ref{det5}). We now use
Eqs.~(\ref{det1})-(\ref{det5}) to prove Eq.~(\ref{detC_hyper1}).
From Eq.~(\ref{det1}), we have
\begin{eqnarray}
\det(C) &=&\prod\limits_{k=1}^{N}\det \left(
\alpha_{kk}^{(N-k)}\right)   \nonumber \\
&=&\det \left( \alpha_{11}^{(N-1)}\right) \ast \det \left( \mathbf{%
\vec{\alpha}}_{22}^{(N-2)}\right) \ast ... \nonumber \\
& \ast & \det \left( \alpha%
_{N-1,N-1}^{(1)}\right) \ast \det \left( \alpha_{NN}^{(0)}\right)
, \label{deter_prove1}
\end{eqnarray}%
where the $\alpha$ matrix stands for
Eqs.~(\ref{det2})-(\ref{det5}) but replacing $A$ matrix for $C$
matrix.

We then apply the same equation for the covariance matrix
$\tilde{C}$
\begin{eqnarray}
\det \tilde{C} &=&\prod\limits_{k=1}^{N}\det \left( \mathbf{\tilde{\alpha}}%
_{kk}^{(N-k)}\right)   \nonumber \\
&=&\det \left( \mathbf{\tilde{\alpha}}_{11}^{(N-1)}\right) \ast
\det \left(
\mathbf{\tilde{\alpha}}_{22}^{(N-2)}\right) \ast ... \nonumber \\
& \ast & \det \left( \mathbf{\tilde{%
\alpha}}_{N-1,N-1}^{(1)}\right) \ast  \det \left( \mathbf{\tilde{\alpha}}%
_{NN}^{(0)}\right) ,  \label{deter_prove2}
\end{eqnarray}%
where the $\tilde{\alpha}$ matrix stands for
Eqs.~(\ref{det2})-(\ref{det5}) but replacing $A$ matrix with
$\tilde{C}$ matrix.

Now we compare the last terms in Eqs.~(\ref{deter_prove1}) and
(\ref{deter_prove2}). Since $\alpha^{0}_{NN}$ is indeed $C_{NN}$
as given by Eq.~(\ref{det2}), we have
\begin{eqnarray}
\det \left(\alpha_{NN}^{(0)}\right)  &=&\det \left( \alpha
_{N}^{-1}\tilde{C}_{NN}\right)   \nonumber \\
&=&\alpha _{N}^{-n_{N}}\det \left( \tilde{C}_{NN}\right)   \nonumber \\
&=&\alpha_{N}^{-n_{N}}\det \left(
\mathbf{\tilde{\alpha}}_{NN}^{(0)}\right).
\end{eqnarray}%
We then calculate the $i$th term; following Eq.~(\ref{det2}), we
have
\begin{eqnarray}
\det \left(\alpha_{ii}^{(N-i)}\right)  &=& C%
_{ii}-\sigma _{i,i+1}\left(C_{N-i}\right) ^{-1}\eta _{i+1,i}
\nonumber \\
&=& C_{ii}-\sum_{m=i+1}^{N}\sum_{n=i+1}^{N} C_{im}\left(C%
\right) _{mn}^{-1}C_{ni}.
\end{eqnarray}%
By using Eq.~(\ref{inverse_new1}), we obtain
\begin{equation}
(C^{-1})_{mn}=(\alpha _{m}\alpha
_{n})^{1/2}(\tilde{C}^{-1})_{mn}\,.
\end{equation}
Therefore we have
\begin{eqnarray}
\det \left(\alpha_{ii}^{(N-i)}\right)  &=&\alpha _{i}^{-1}%
\tilde{C}_{ii}-\sum_{m=i+1}^{N}\sum_{n=i+1}^{N}(\alpha _{m}\alpha
_{i})^{-1/2} \nonumber \\
& \times &
 \tilde{C}_{im}(\alpha _{m}\alpha _{n})^{1/2}\left(C
\right)
_{mn}^{-1}(\alpha _{i}\alpha _{n})^{-1/2} C_{ni}  \nonumber \\
&=&\alpha _{i}^{-1}\left(\tilde{C}_{ii}-\sum_{m=i+1}^{N}\sum_{n=i+1}^{N}%
\tilde{C}_{im}\left( \tilde{C}\right)
_{mn}^{-1}\tilde{C}_{ni}\right)
\nonumber \\
&=&\alpha _{i}^{-1}\det \left(
\mathbf{\tilde{\alpha}}_{ii}^{(N-i)}\right) .
\label{eq:proof2-last}
\end{eqnarray}%
Thus, by mathematical induction, we have proved that all of the
terms in Eqs.~(\ref{deter_prove1}) and (\ref{deter_prove2})
follow Eq.~(\ref{eq:proof2-last}). Therefore the relationship between Eqs.~(\ref{deter_prove1}) and (\ref%
{deter_prove2}) is
\begin{equation}
\det (C)=\det (\tilde{C}) \ast \left( \prod\limits_{i=1}^{N}\alpha
_{i}^{-n_{i}}\right) ,
\end{equation}%
i.e. we have proved Eq.~(\ref{detC_hyper1}). $\Box$

Combining Proofs (1) and (2), we have shown that, in general, when
combining multiple correlated data sets with hyperparameters, the
inverse and determinant of the covariance matrix follow
Eqs.~(\ref{inverse_new1}) and (\ref{detC_hyper1}). Therefore the
likelihood function for combined correlated data sets is
Eq.~(\ref{like4}).

Equation~(\ref{like4}) greatly simplifies the computation of
hyperparameter likelihood, since one can always calculate the
covariance matrix for correlated data sets $\tilde{C}$ and then
use ``element-wise'' product $\odot$ to calculate the covariance
matrix with hyperparameters, and then numerically solve for the
maximum likelihood solution.

\section{Hadamard product and inverse}
\label{app:hadamard}

The Hadamard product is the element-wise product of any two
matrices with the same dimension. If $A$ and $B$ are the two
matrices with the same dimension $m \times n$, the Hadamard
product $A\circ B$ is a matrix with the same dimension with
element ($i,j$) equal to
\begin{eqnarray}
(A \circ B)_{ij}=A_{ij}\cdot B_{ij} .\label{eq:hadamard}
\end{eqnarray}

The Hadamard inverse is an inverse operation which requires that
each element of the matrix is nonzero, so that each element of the
Hadamard inverse matrix is
\begin{equation}
\hat{A}_{ij}=A^{-1}_{ij}. \label{eq:hadamard-inverse}
\end{equation}%
Here we use a hat to denote the Hadamard inverse. Therefore the
Hadamard product of an $m \times n$ matrix and its Hadamard
inverse becomes a unit matrix where all elements are equal to one,
i.e.
\begin{equation}
A \circ \hat{A}=(J)_{m \times n}. \label{eq:hadamard-product1}
\end{equation}%

\section{A lemma for determinant}
\label{block_matrix2}

We will use the following Lemma to prove the determinant relation
of the covariance matrix of hyperparameter likelihood,
Eq.~(\ref{detC_hyper1}).

Let $A$ be an ($N_{t}\times N_{t})$ real or complex matrix, which
is
partitioned into $N\times N$ blocks, each of size is $n_{i}\times n_{j},$%
which satisfies
\begin{equation}
\sum_{i=1}^{N}n_{i}=N_{t}.
\end{equation}%
\begin{equation}
A=\left(
\begin{array}{cccc}
\left( A_{11}\right) _{n_{1}\times n_{1}} & \left( A_{12}\right)
_{n_{1}\times n_{2}} & ... & \left( A_{1N}\right) _{n_{1}\times n_{N}} \\
(A_{12})_{n_{2}\times n_{1}}^{T} & \left( A_{22}\right)
_{n_{2}\times n_{2}}
& ... & \left( A_{2N}\right) _{n_{2}\times n_{N}} \\
... & ... & ... & ... \\
(A_{1N})_{n_{N}\times n_{1}}^{T} & (A_{2N})_{n_{N}\times
n_{2}}^{T} & ... &
\left( A_{NN}\right) _{n_{N}\times n_{N}}%
\end{array}%
\right).  \label{A_mat}
\end{equation}%
The determinant of $A$ is given by
\begin{equation}
\det A\mathbf{=}\prod\limits_{k=1}^{N}\det \left( \mathbf{\alpha }%
_{kk}^{(N-k)}\right) ,  \label{det1}
\end{equation}%
where $\mathbf{\alpha }^{(k)}$ is defined as
\begin{eqnarray}
\mathbf{\alpha }_{ij}^{(0)} &=&A_{ij}  \nonumber \\
\mathbf{\alpha }_{ij}^{(k)} &=&A_{ij}-\sigma _{i,N-k+1}\left( \bar{A}%
_{k}\right) ^{-1}\eta _{N-k+1,j},(k\geqslant 1),  \label{det2}
\end{eqnarray}%
where vectors $\sigma _{ij}^{T}$ and $\eta _{ij}$ are defined as
\begin{equation}
\sigma_{ij}=\left( A_{ij},\textrm{ }A_{i,j+1},...\textrm{
}A_{i,N}\right) , \label{det3}
\end{equation}%
\begin{equation}
\eta _{ij}=\left( A_{ij}, \textrm{ }A_{i+1,j},...\textrm{
}A_{N,j}\right) ^{T}, \label{det4}
\end{equation}%
and $\bar{A}_{k}$ is defined as
\begin{equation}
\bar{A}_{k}=\left(
\begin{array}{cccc}
A_{N-k+1,N-k+1} & A_{N-k+1,N-k+2} & ... & A_{N-k+1,N} \\
A_{N-k+2,N-k+1} & A_{N-k+2,N-k+2} & ... & A_{N-k+2,N} \\
... & ... & ... & ... \\
A_{N,N-k+1} & A_{N,N-k+2} & ... & A_{N,N}%
\end{array}%
\right) .  \label{det5}
\end{equation}

A particular case of this lemma, where each block matrix has the
same dimension $n \times n$, is shown as a theorem in
\cite{Powell11}. Here we extend the theorem shown in
\citet{Powell11} to a more general case where each diagonal block
matrix may have a different size, so the off-diagonal matrix can
be a rectangular matrix.

\textbf{Proof.} We start from the simplest case, where $N=2$, i.e. $A$ is a $%
2\times 2$ symmetric block matrix
\begin{equation}
A=\left(
\begin{array}{cc}
A_{11} & A_{12} \\
A_{12}^{T} & A_{22}%
\end{array}%
\right) ,
\end{equation}%
where $A_{11}$ and $A_{22}$ are $p\times p$ and $q\times q$
semi-positive definite symmetric matrix respectively, and $A_{12}$
is a $p\times q$ matrix. The determinant of $A$ is
\begin{eqnarray}
\det A &=&\det \left( A_{11}-A_{12}A_{22}^{-1}A_{12}^{T}\right)
\det \left(
A_{22}\right)  \nonumber \\
&=&\det \left( A_{22}-A_{12}A_{11}^{-1}A_{12}^{T}\right) \det
\left( A_{11}\right) .  \label{det2by2}
\end{eqnarray}%
We can immediately check that this is indeed the simplest case for Eqs.~(\ref%
{det1})-(\ref{det5}) where $N=2$. Since if $N=2$, Eq.~(\ref{det2})
gives $\det A\mathbf{=}\det \left( \mathbf{\alpha
}_{11}^{(1)}\right) \det \left(
\mathbf{\alpha }_{22}^{(0)}\right) ,$ where $\mathbf{\alpha }%
_{22}^{(0)}=A_{22},$ and $\mathbf{\alpha }%
_{11}^{(1)}=A_{11}-A_{12}A_{22}^{-1}A_{12}^{T}$, which is exactly
Eq.~(\ref{det2by2}).

Now we can use Eq.~(\ref{det2by2}) for the $N=2$ case to inductively derive general equations (\ref{det1})-(\ref%
{det5}). Let us treat matrix (\ref{A_mat}) as a 2-by-2 matrix,
where all of the matrices $A_{22,}A_{33},...A_{NN}$ are grouped into a big matrix $%
\mathbf{\tilde{A}}_{22}:$
\begin{equation}
A=\left(
\begin{array}{cc}
A_{11} & \mathbf{\tilde{A}}_{12} \\
\mathbf{\tilde{A}}_{12}^{T} & \mathbf{\tilde{A}}_{22}%
\end{array}%
\right) ,
\end{equation}%
where
\begin{equation}
\mathbf{\tilde{A}}_{22}=.\left(
\begin{array}{cccc}
A_{22} & A_{23} & ... & A_{2N} \\
A_{23}^{T} & A_{33} & ... & A_{3N} \\
... & ... & ... & ... \\
A_{2N}^{T} & A_{3N}^{T} & ... & A_{NN}%
\end{array}%
\right) ,
\end{equation}%
is exactly $\bar{A}_{N-1}$ (Eq.~(\ref{det5})), and
\begin{equation}
\mathbf{\tilde{A}}_{12}=\left(
\begin{array}{cccc}
A_{12} & A_{13} & ... & A_{1N}%
\end{array}%
\right) ,
\end{equation}%
is exactly the definition of $\sigma_{12}$ (Eq.~(\ref{det3})). In
addition,
\begin{equation}
\mathbf{\tilde{A}}_{12}^{T}=\left(
\begin{array}{cccc}
A_{21} & A_{31} & ... & A_{N1}%
\end{array}%
\right) ^{T},
\end{equation}%
is exactly $\eta_{21}$ (Eq.~(\ref{det4})). Now applying the second line of Eq.~(%
\ref{det2by2}) to this matrix, one has
\begin{eqnarray}
\det \left( A\right) &=&\det \left( \mathbf{\tilde{A}}_{22}\right)
\ast \det
\left( A_{11}-\mathbf{\tilde{A}}_{12}\mathbf{\tilde{A}}_{22}^{-1}\mathbf{%
\tilde{A}}_{12}^{T}\right)  \nonumber \\
&=&\det \left( \mathbf{\tilde{A}}_{22}\right) \ast \det \left(
A_{11}-\sigma _{12}\bar{A}_{N-1}^{-1}\eta _{21}\right) .
\label{detA221}
\end{eqnarray}%
Now proceeding to $\det \left( \mathbf{\tilde{A}}_{22}\right) ,$ again, $%
\mathbf{\tilde{A}}_{22}$ can be separated into two big matrices as%
\begin{equation}
\mathbf{\tilde{A}}_{22}=\left(
\begin{array}{cc}
A_{22} & \mathbf{\tilde{A}}_{23} \\
\mathbf{\tilde{A}}_{23}^{T} & \mathbf{\tilde{A}}_{33}%
\end{array}%
\right) ,
\end{equation}%
where
\begin{equation}
\mathbf{\tilde{A}}_{33}=.\left(
\begin{array}{cccc}
A_{33} & A_{34} & ... & A_{3N} \\
A_{34}^{T} & A_{44} & ... & A_{4N} \\
... & ... & ... & ... \\
A_{3N}^{T} & A_{4N}^{T} & ... & A_{NN}%
\end{array}%
\right) =\bar{A}_{N-2},
\end{equation}%
and
\begin{eqnarray}
\mathbf{\tilde{A}}_{23} &=&\left(
\begin{array}{cccc}
A_{23} & A_{24} & ... & A_{2N}%
\end{array}%
\right) =\sigma _{23},  \nonumber \\
\mathbf{\tilde{A}}_{23}^{T} &=&\left(
\begin{array}{cccc}
A_{32} & A_{42} & ... & A_{N2}%
\end{array}%
\right) ^{T}=\eta _{32},
\end{eqnarray}%
therefore
\begin{eqnarray}
\det \left( \mathbf{\tilde{A}}_{22}\right) &=&\det \left( \mathbf{\tilde{A}}%
_{33}\right) \ast \det \left( A_{22}-\mathbf{\tilde{A}}_{23}\mathbf{\tilde{A}%
}_{33}^{-1}\mathbf{\tilde{A}}_{23}^{T}\right)  \nonumber \\
&=&\det \left( \mathbf{\tilde{A}}_{33}\right) \ast \det \left(
A_{22}-\sigma _{23}\bar{A}_{N-2}^{-1}\eta _{32}\right) ,
\label{detA22}
\end{eqnarray}%
so combining Eqs.~(\ref{detA22}) and (\ref{detA221}), we have
\begin{eqnarray}
\det \left( A\right) &=& \det \left(
\mathbf{\tilde{A}}_{33}\right) \ast \det \left( A_{22}-\sigma
_{23}\bar{A}_{N-2}^{-1}\eta
_{32}\right) \nonumber \\
 &\ast& \det \left( A_{11}-\sigma
_{12}\bar{A}_{N-1}^{-1}\eta _{21}\right) .
\end{eqnarray}%
Repeating this operation until breaking down the first term, one can eventually reach $%
A_{NN},$ therefore the determinant of $A$ is
\begin{eqnarray}
\det \left( A\right) &=&\det \left( A_{NN}\right) \nonumber \\
& \ast & \det \left(
A_{N-1,N-1}-A_{N-1,N}\bar{A}_{1}^{-1}A_{N,N-1}\right) \nonumber \\
& \ast & ...\ast \det \left( A_{22}-\sigma
_{23}\bar{A}_{N-2}^{-1}\eta _{32}\right) \nonumber \\
 &\ast& \det
\left( A_{11}-\sigma _{12}\bar{A}_{N-1}^{-1}\eta _{21}\right)
\label{eq:detA1}.
\end{eqnarray}%
By comparing the brackets in Eq.~(\ref{eq:detA1}) with
Eq.~(\ref{det2}), one can find that each term is exactly the same,
therefore the determinant is given by Eq.~(\ref%
{det1}). $\Box$

\end{document}